\pdfoutput=1
\documentclass[conference]{IEEEtran}
\IEEEoverridecommandlockouts
\usepackage{cite}
\usepackage{amsmath,amssymb,amsfonts}
\usepackage{graphicx}
\usepackage{textcomp}
\usepackage{xcolor}
\def\BibTeX{{\rm B\kern-.05em{\sc i\kern-.025em b}\kern-.08em
    T\kern-.1667em\lower.7ex\hbox{E}\kern-.125emX}}

\usepackage{subfig}
\usepackage{graphicx}
\usepackage{graphics}
\usepackage{epstopdf}
\usepackage{amsmath}
\usepackage{multirow}
\usepackage{bm}
\usepackage{enumerate}
\usepackage{mathdots}
\usepackage{todonotes}
\usepackage{algorithmicx}
\usepackage{algpseudocode}
\usepackage{algorithm}
\usepackage{verbatim}
\usepackage{balance}
\usepackage{cite}
\usepackage{url}
\usepackage{color}
\definecolor{purple(x11)}{rgb}{0.63, 0.36, 0.94}
\definecolor{cadmiumgreen}{rgb}{0.0, 0.42, 0.24}

\graphicspath{{./images/}}

\begin{document}
\title{A Dynamic Codebook Design for Analog Beamforming in MIMO LEO Satellite Communications}
\author{Joan Palacios\,$^{1}$, Nuria Gonz\'{a}lez-Prelcic\,$^{1}$, Carlos Mosquera\,$^{2}$ and Takayuki Shimizu \,$^{3}$\\
  $^{1}$Electrical and Computing Engineering Department, North Carolina State University,  Raleigh, NC, USA \\
  $^{2}$atlanTTic Research Center, Universidade de Vigo, Vigo, Spain\\
  $^{3}$Toyota Motor North America, Mountain View, CA, USA
}

\maketitle

\begin{abstract}
Beamforming gain is a key ingredient in the performance of LEO satellite communication systems to be integrated into cellular networks.
However, beam codebooks previously designed in the context of MIMO communication for terrestrial networks, do not provide the appropriate performance in terms of inter-beam interference and gain stability as the satellite moves. In this paper, we propose a dynamic codebook that provides a stable gain during the period of time that the satellite covers a given cell, while avoiding link retraining and extra calculation as the  satellite moves. In addition, the proposed codebook provides a higher signal-to-interference-plus-noise (SINR) ratio than those DFT codebooks commonly used in cellular systems.

\end{abstract}

%
\section{Introduction}
New flexible payloads are becoming more common in emerging satellite services due to significant  innovations in active antennas and modular processing blocks, among others \cite{riccardo2020}, \cite{Angeletti2020}. These new technologies pave the way for the integration of satellite communications in general, and LEO constellations in particular,  into future cellular networks. However, many open challenges still need to be addressed at the physical layer, including the design of appropriate beamforming strategies, modulation and coding schemes or approaches for link adaption \cite{Kodheli2017}.

To provide the required beamforning gain, analog fully reconfigurable beamforming networks entail high mass and power dissipation, limiting the number of beams. In the particular case of LEO satellites, large efforts are underway to develop hybrid arrays combining analog beamforming at a subarray level with digital processing of subarrays.  Despite the expected performance, especially for large arrays, the flexibility comes at a cost, in such a way that the most relevant massive LEO constellations \cite{delPortillo2019,Xia2019}  which are either  operational or near to be, still make use of fixed analog beams.  However, and similarly to terrestrial cellular systems, some degree of beam steering is desirable to focus the resources at will, keeping in mind that the size of satellite beams is much larger than that for terrestrial beams, able to offer better spatial discrimination. 
 
Beam codebooks for cellular networks being deployed are based on an oversampled  2-D Discrete Fourier Transform (DFT) type grid of beams  \cite{Miao2018,R1-1709232}. This design was adapted in \cite{PalaciosLEO21} to LEO systems by using the interpolation factor to adjust the beamwidth to the size of  the region of interest. However, as shown in \cite{PalaciosLEO21}, using this type of codebook results in a high signal-to-interference-plus-noise ratio (SINR) and a large variation in beamforming gain  inside the beam switching time. In addition, the handovers between beams when using a static codebook are very frequent, degrading the user experience \cite{Su2019}.  New designs that account for the mobility of the LEO satellite  and  the short amount of time that a given location is under the coverage of a LEO beam, need to be devised. 

In this paper we address the problem of analog beamforming for LEO payloads with limited reconfigurability, by designing a family of codebooks that, together with the appropriate switching strategy, can steer the beams with limited resolution on the region to serve. The proposed design minimizes the number of handovers between beams for the ground terminal while keeping a stable beamforming gain during the period of time that the same LEO satellite is serving a given user.

\section{System model}

We consider the link between a LEO satellite and a user terminal operating in the Ku band with a bandwidth $B$ and a full frequency reuse (FFR) scheme across beams. 
The user terminal is located in an elliptical region of interest (ROI) or coverage area with semi-radius $R_{\rm x}$ and $R_{\rm y}$, designed following a given criteria, for example the one described in \cite{PalaciosLEO21}.
We denote with ${\rm x}$ the axis in the direction of the satellite's movement, while ${\rm y}$ is the axis orthogonal to ${\rm x}$.
The satellite is operating at a height $h_{\rm sat}$ with an associated
angular speed of $w_{\rm sat} = \frac{v_{\rm sat}}{r_{\rm earth}+h_{\rm sat}}$, where $R_{\rm earth}$ is the Earth's radius and 
$v_{\rm sat}$ is the satellite linear speed.
Note that the satellite linear speed also depends on the orbit height as
\begin{equation}
v_{\rm sat} = \sqrt{G\frac{m_{\rm earth}}{R_{\rm earth}+h_{\rm sat}}},
\end{equation}
where $m_{\rm earth}$ is the Earth's mass, and $G$ is the gravitational constant. We also define the region of proximity (ROP) as the elliptical region of the same size as the ROI located right under the satellite. Note that the ROI and ROP  do not have to be exactly the same, since the exact location of the ROI depends on the main directions of the beampatterns generated at the satellite.

The satellite is equipped with $N_{\rm RF}$ uniform planar sub-arrays of size $N^\text{sub}_{\rm x}\times N^\text{sub}_{\rm y}$, for a total number of antenna elements $N^{\rm sat} = N_{\rm RF}N^\text{sub}_{\rm x}N^\text{sub}_{\rm y}$.
Note that while it is common to distribute the sub-arrays in a rectangular way, since each sub-array will be working independently of the others, their particular distribution will be irrelevant. By having the relative antenna elements positions defined by ${\bf k}_{\rm sat}\in\mathbb{R}^{N^{\rm sat}\times3}$,
with each of the radiating elements position at $[{\bf k}_{\rm sat}]_{n, :}$, and denoting as $\mathcal{S}_2=\{{\bf x}\in\mathbb{R}^3\text{ s.t. }\|{\bf x}\|=1\}$ the unitary sphere, the steering vectors that describe the response of the satellite's antenna in the direction ${\bf v}\in\mathcal{S}_2$ are denoted as ${\bf a}_{\rm sat}: \mathcal{S}_2\rightarrow\mathbb{C}^{N^{\rm sat}}$ and defined as
\begin{equation}\label{eq:steering_sat}
{[{\bf a}_{\rm sat}({\bf v})]_n} = e^{-j\frac{2\pi}{\lambda}<{[\bf k}_{\rm sat}]_{n, :}, {\bf v}>}.
\end{equation}
   
The LEO satellite serves a number $N_\text{u}$ of mobile user terminals (UTs) on  the ground by illuminating the ROI with $N_{\rm b}$ analog beams that are chosen from a given codebook and move with the satellite. 
We denote the analog precoding matrix used by the satellite  as ${\bf F}_{\rm RF}\in\mathbb{C}^{N^\text{sat} \times N_{\rm b}}$. Given that the analog precoding stage will be implemented by means of a phase shifting network, we can write
\begin{equation}
[{\bf F}_{\rm RF}]_{n, m} = \left\{\begin{array}{rl}
    e^{j\phi_n} & {\rm if } \quad m = I_{\rm RF}(n) \\
    0 & {\rm otherwise},
\end{array}\right.
\end{equation}
where $I_{\rm RF}(n)$ is the index of the RF-chain to which the $n$-th antenna is connected.
As in \cite{Kim2020}, we assume that the UTs exploit satellite's position information for beam tracking. 
We are considering a uniform planar array (UPA) of size $N^{\rm UT}=N^\text{UT}_{\rm x}\times N^\text{UT}_{\rm y}$ at the terminals. 
The UT steering vector is denoted as ${\bf a}_{\rm u}:\mathcal{S}_2\rightarrow\mathbb{C}^{N_{\rm u}}$, and is defined as ${[{\bf a}_{\rm u}({\bf v})]_n} = e^{-i\frac{2\pi}{\lambda}<{[\bf k}_{\rm u}]_{n, :}, {\bf v}>}$, where ${\bf k}_{\rm u}\in\mathbb{C}^{N_{\rm u}\times 3}$ are the antenna elements relative positions.
The UT combiner is denoted as ${\bf w}\in\mathbb{C}^{N^{\rm UT}}$, and is assumed to be analog-only.

We adopt a Rician  geometric channel model with LoS complex gain $\gamma$ and Rician factor $K_{\rm r}$ as in \cite{PalaciosLEO21}.
Given the steering vectors for the satellite and the UT previously defined, and assuming perfect symbol and carrier synchronization, the equivalent channel matrix is 
\begin{equation}\label{eq:rician_channel}
  {\bf H} = \gamma\left({\bf a}_{\rm u}\left(\frac{{\bf x}_{\rm u}-{\bf x}_{\rm sat}}{\|{\bf x}_{\rm u}-{\bf x}_{\rm sat}\|}\right)
    +\sqrt{\frac{1}{K_{\rm r}}}{\bf a}_{\rm R}\right){\bf a}_{\rm sat}^{\rm H}\left(\frac{{\bf x}_{\rm sat}-{\bf x}_{\rm u}}{\|{\bf x}_{\rm sat}-{\bf x}_{\rm u}\|}\right)
\end{equation}
where ${\bf x}_{\rm sat}$ is the satellite position, ${\bf x}_{\rm u}$ is the UT position, and ${\bf a}_{\rm R}\sim\mathcal{CN}(0, \bf  \Sigma)$ is the Rician component satisfying
${\rm trace}(\bf \Sigma) = \|{\bf a}_{\rm u}\left(\frac{{\bf x}_{\rm u}-{\bf x}_{\rm sat}}{\|{\bf x}_{\rm u}-{\bf x}_{\rm sat}\|}\right)\|^2$.
The term $\gamma$ includes all path loss effects, dominated by the free space path loss $LP_{\rm fs}{\rm [dB]}$ and the atmospheric attenuation $LP_{\rm at}{\rm [dB]}$.

 \begin{figure}[t!]
    \centering
        \vspace*{2mm}
    \includegraphics[width = 0.95\linewidth]{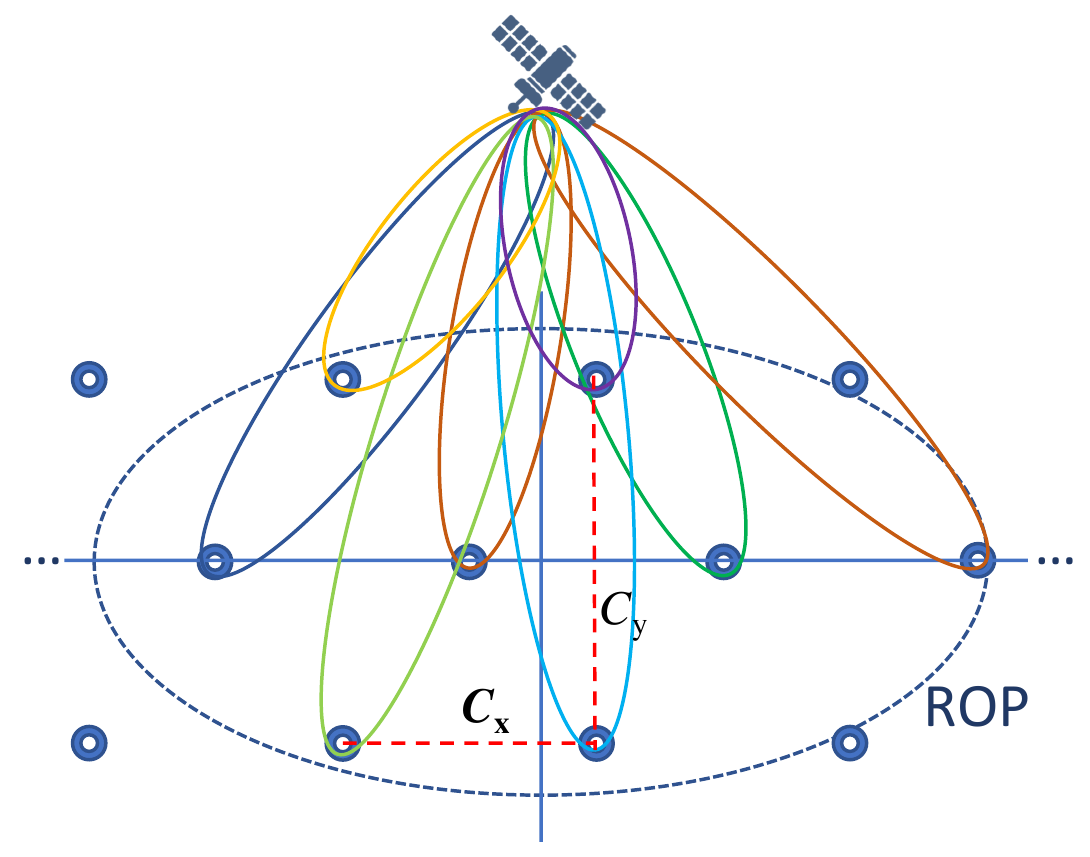}
    \caption{Illustration of the idea behind the design of the initial codebook. The beams in the initial codebook are designed to have maximum gain at the points in the hexagonal lattice which fall inside the satellite's ROP. }
        \label{fig:initial_codebook}
\end{figure}

The signal-to-noise ratio (SNR) can be written in terms of the received signal strength (RSS) and the noise power $\sigma^2$ as \cite{TR38.821} 
\begin{align}
SNR{\rm [dB]}=RSS{\rm [dBW]}-\sigma^2{\rm [dBW]}\\
RSS{\rm [dBW]}=P_{\rm TX}{\rm [dBW]}-LP_{\rm cable}{\rm [dB]}+G_{\rm TX}{\rm [dB]} \nonumber \\
-LP_{\rm at}{\rm [dB]}-LP_{\rm fs}{\rm [dB]}+G_{\rm RX}{\rm [dB]}\\
\sigma^2{\rm [dBW]}=T{\rm [dBK]}+k{\rm [dBW/K/Hz]}+B{\rm [dBHz]},
\end{align}
where $P_{\rm TX}$ is the transmit power, $G_{\rm TX}$ is the transmit antenna gain, $LP_{\rm cable}$ is the cable loss between the antenna and the transmitter, and $G_{\rm RX}$ is the receiver antenna gain. Additionally,  ${T}$ and $k$ denote  the noise temperature and the Boltzmann constant, respectively. 
The receiver gain $G_{\rm TX}$ of this antenna when considering a Rician channel is extracted from \cite{PalaciosLEO21} to be
\begin{equation}
G_{\rm RX} = 10\log_{10}\left(N^\text{UT}_{\rm x}N^\text{UT}_{\rm y}+\frac{1}{K_{\rm r}}\right).
\end{equation}
Same as in \cite{PalaciosLEO21}, we include the gain due to capturing  the Rician component of the channel in the receiver as a natural result to ease the formulation.

\section{Dynamic codebook design} 
\label{sec:analog_design}
Given a specific coverage area, our goal is to design a codebook of analog precoders  that provides appropriate coverage, minimizing the SNR and SINR loss due to the satellite movement and the static beams. This loss has been described and numerically evaluated in  \cite{PalaciosLEO21} for a 2D DFT beam codebook as that used in 5G New Radio (NR). In this section, we design a set of feasible analog precoders that enable seamless communication without having to re-train the link every time a user is not covered anymore by a given beam due to the satellite's  movement.

There are three main ideas behind our design: 1) selection of a basic codebook that provides a better SINR than that associated to the DFT codebook; 2) introduction of a codebook adaptation mechanism so the SNR variation due to the satellite movement is compensated; and 3) definition of a beam ID permutation protocol over the codebook that minimizes handovers. In the next paragraphs we describe these different components of the final design.

  \begin{figure}[t!]
    \centering
     \vspace*{2mm}
    \includegraphics[width = 0.95\linewidth]{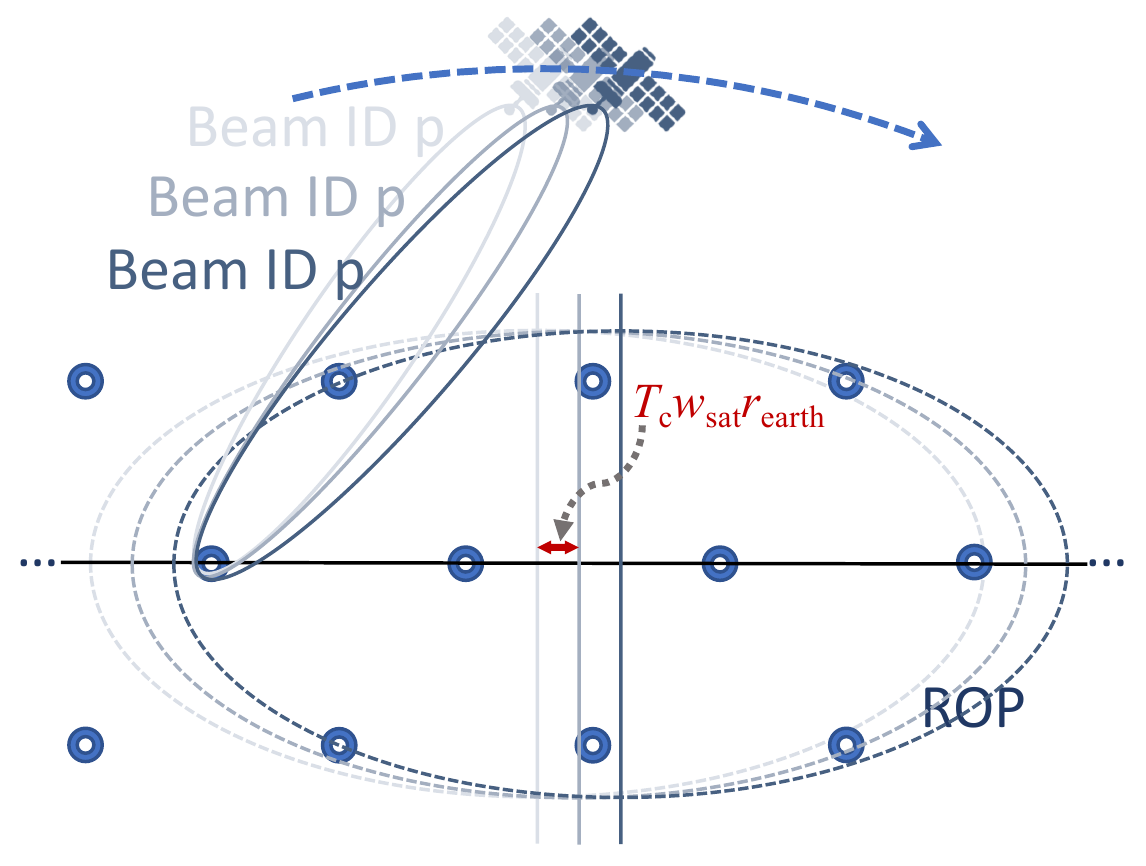}
    \caption{Illustration of the idea behind the dynamic codebook. The beams in the codebook are redesigned every $T_{\rm c}$ to compensate for the satellite's movement. The number of active beams corresponds to the number of points in the triangular lattice that fall inside the satellite's ROP every time the beam codebook is updated.}
        \label{fig:dynamic_codebook}
\end{figure}

\subsection{Initial codebook design}\label{sec:initial_codebook}
For the initial codebook design we choose to create beams whose maximum gains appear at the points of a given triangular (or hexagonal) lattice, as illustrated in Fig. \ref{fig:initial_codebook}. The  points in the triangular lattice are scaled to be proportional to the diameter covered by a beam-width by a factor $C_{\rm x} = \frac{\pi h_{\rm sat}}{ON^{\rm sub}_{\rm x}}$ in the ${\rm x}$ direction
and  by a factor $C_{\rm y} = \frac{\pi h_{\rm sat}}{ON^{\rm sub}_{\rm y}}$ in the ${\rm x}$ direction, where $O$ is defined as the oversampling factor. 
The value of the oversampling factor is chosen to adjust the number of points of the triangular lattice that will fall inside the satellite's shadow, i.e., the number of beams that will illuminate this region. The choice of a triangular lattice over a rectangular one (as that associated to a DFT codebook), gives rise to the common hexagonal tesselation in satellite footprints, and prevents the existence of locations with up to four beams with equal gain, highly detrimental for the signal-to-interference-plus-noise ratio (SINR).
In addition, the alignment of the x-axis with the triangulation minimizes the periodicity interval of the lattice in that direction.

\begin{figure*}[t!]
    \centering
    \vspace*{9mm}
    \includegraphics[width = 0.58\linewidth]{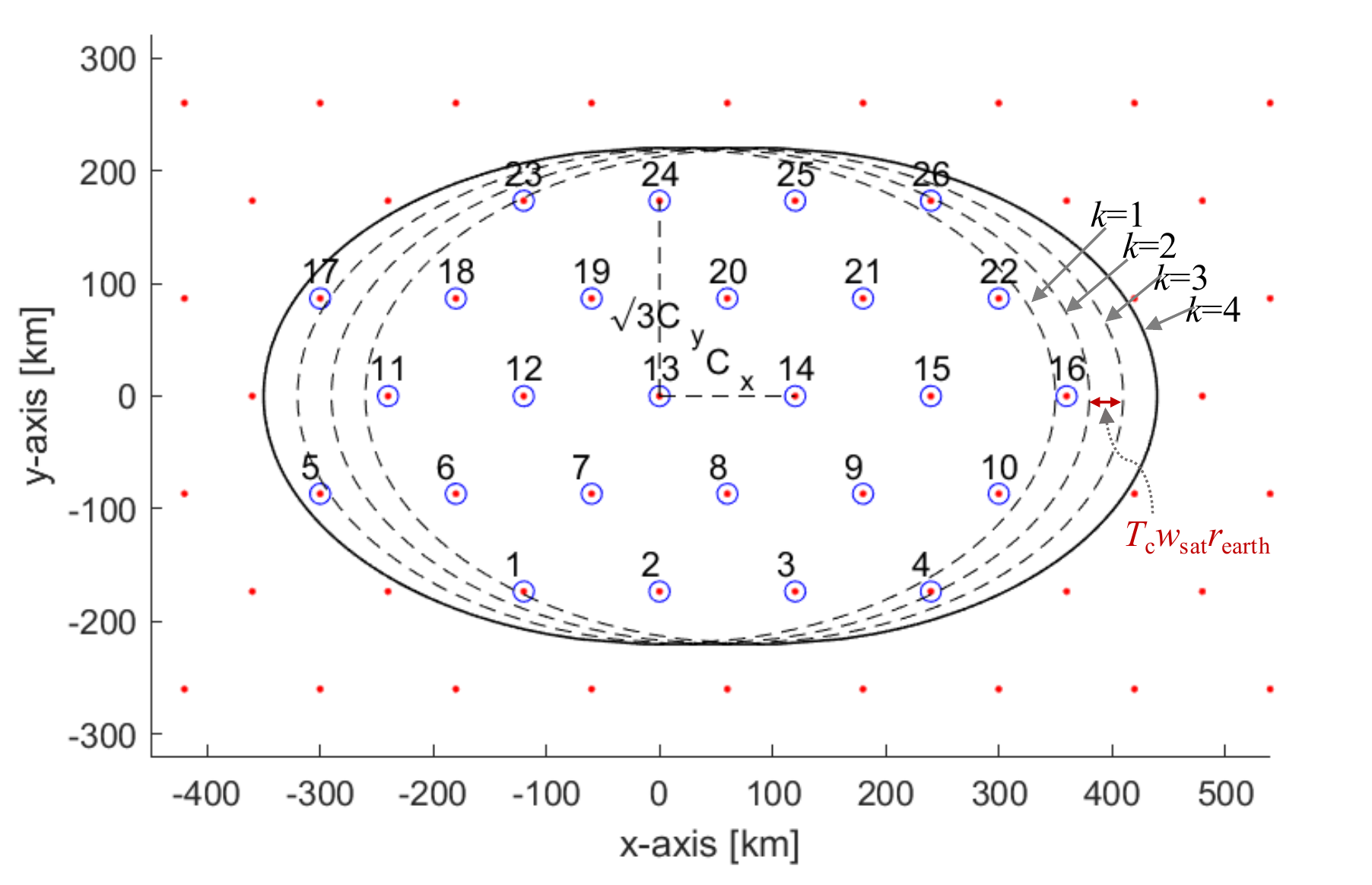}
    \caption{Representation of the hexagonal lattice  $\mathcal{L}_k$, for a toy example with $K=4$. The points of the hexagonal lattice are marked in red. Each ellipse represents the ROP for different time lapses.  The beams IDs are updated so there is no need of handover at the user terminal, which always sees the same ID over the time the same LEO satellite is illuminating the cell. The  subset of points that are or will be active in any of the lattice translations are circled in blue, and labeled following the proposed criteria.}
        \label{fig:hex}
\end{figure*}

\subsection{Codebook adaptation}
 The main idea behind our design consists of automatically updating the analog precoder every time interval of length  $T_c$  to compensate for the satellite's movement without the explicit need of computing a new precoder, but simply playing it from a dynamic codebook in a predefined order. This is illustrated in Fig.~\ref{fig:dynamic_codebook} for one of the beams in the initial codebook. This way, the beam pattern that points to a specific point in the triangular lattice is updated in a temporal grid to keep a high gain while that point is within the reach of the satellite's ROP. In an ideal system, the beampattern  would be adapted instantaneously as the satellite moves, but in a practical system only a periodic adaptation is feasible.

 To simplify the  dynamic codebook design problem, we adopt the geometry of the stereographic projection (ignoring the Earth's curvature and assuming the origin of coordinates is at the satellite's location),  and using the unitary direction of the satellite movement as x-axis and the perpendicular to the satellite movement and the orbital plan as the y-axis. The inaccuracies  generated by this adoption can be neglected  in the context of a LEO constellation, due to the high altitude of the satellites.
Under this assumption, the condition of maximizing the stability of the directionality over the Earth's surface is equivalent to correcting the relative movement of the UT with  respect to the satellite due to the satellite  movement, i.e.  ${\bf v}_{\rm u}' = [-r_{\rm Earth}w_{\rm sat}, 0, 0]$.
Note that during the time interval $T_{\rm c}$ the point of maximum directionality relative to the satellite moves a distance of $T_{\rm c}r_{\rm Earth}w_{\rm sat}$ in the opposite direction to the satellite's movement. This tells us how to design the subsequent codebooks given an initial codebook. 

Now we define the spatial shift in the $\rm x$ axis of the satellite's ROP over a period $KT_{\rm c}$ as $\Delta=KT_{\rm c}w_{\rm sat}r_{\rm earth}$ for a given integer number $K$. For an efficient implementation of  the dynamic codebook, we realize that by setting $T_{\rm c}$ such that the scaling factor satisfies $C_{\rm x}=\Delta$, the lattice becomes invariant to the spatial shift $\Delta$.
This means that by selecting an integer number $K$, we can create a cyclic dynamic codebook where each iteration lasts $T_{\rm c}$, repeats every $K$ iterations (cycle), and, as we will explain in Section~\ref{sec:permutation}, shuffles indices for coherence.
Mathematically, the lattice points of the $k$-th iteration can be expressed as
\begin{equation}
\begin{array}{rl}
\mathcal{L}_k = & C_{\rm x}(-\frac{k}{K}+\mathbb{Z})\times\sqrt{3}C_{\rm y}\mathbb{Z}\\
 & \bigcup C_{\rm x}(\frac{1}{2}-\frac{k}{K}+\mathbb{Z})\times\sqrt{3}C_{\rm y}(\frac{1}{2}+\mathbb{Z}),\\
 T_{\rm c} = & \frac{\pi h_{\rm sat}}{Kw_{\rm sat}r_{\rm earth}oN^{\rm sub}_{\rm x}}.
\end{array}
\end{equation}
$\mathcal{L}_k$ describes the lattice points assuming an infinite lattice, although for the definition of  the final codebook we just need to keep those points that fall inside the 
coverage area. The condition for a point $(x,y)$ being inside the elliptical ROI $\mathcal{E}$ is $(\frac{x}{R_{\rm x}})^2+(\frac{y}{R_{\rm y}})^2 \leq 1$. Thus, we select the points in $\mathcal{L}_k$ that fulfill this equation.
We do this for each iteration $k$, leading to a different number of points being inside the ellipse over time as $k$ changes, and in consequence having a slightly different number of RF-chains and active beams in use over time. This process is illustrated in Fig.~\ref{fig:hex} for $K=4$, i.e., a dynamic codebook built from 4 static codebooks in 4 iterations. The duration if the cycle is $4T_{\rm c}$ in this case. Note that in each iteration, the number of active beams can decrease, increase, or keep unchanged, depending on the set of points of the triangular lattice that stay inside the corresponding ROP. 

To compute the expression for the beamformers in the dynamic codebook as a function of the iteration $k$, by considering a point $(x, y)$ at height $0$ with origin on the satellite?s shadow, the direction from the satellite to that point can be computed as $\frac{{\bf x}_{\rm u}-{\bf x}_{\rm sat}}{\|{\bf x}_{\rm u}-{\bf x}_{\rm sat}\|} = \frac{1}{\sqrt{x^2+y^2+h_{\rm sat}^2}}(x, y, -h_{\rm sat})$. Constructive interference to create a narrow beam-pattern in that direction can be achieved by setting 
the satellite's $n$-th antenna's phase-shift equal to the normalized satellite's steering vector coefficient evaluated at that direction, i.e., $[{\bf a}_{\rm sat}^{\rm H}\left(\frac{{\bf x}_{\rm sat}-{\bf x}_{\rm u}}{\|{\bf x}_{\rm sat}-{\bf x}_{\rm u}\|}\right)]_n/\sqrt{N_{\rm x}^{\rm sub}N_{\rm y}^{\rm sub}}$. 
Note that the normalization factor depends on the size of the subarray connected to a specific RF chain.
In consequence, for every iteration $k$, with one RF-chain per beam pointing to $(x_m, y_m)\in\mathcal{L}_k\bigcap\mathcal{E}$,  we can set the phase-shift coefficient of the $n$-th antenna connected to that RF-chain to 
\begin{equation}
[{\bf F}_{{\rm RF}, k}]_{n, m} =
\frac{e^{-j\pi\frac{1}{\sqrt{x_m^2+y_m^2+h_{\rm sat}^2}}(x_m[{\bf k}]_{n, 1} + y_m[{\bf k}]_{n, 2})}}{\sqrt{N_{\rm x}^{\rm sub}N_{\rm y}^{\rm sub}}},
\end{equation}
for $I_\text{RF}(n)=m$, and $0$ otherwise.

\subsection{Beam IDs permutations}\label{sec:permutation}
Beam IDs are permuted at each iteration if the dynamic codebook to guarantee that the beam ID covering a given hexagonal area is static, as illustrated in Fig.~\ref{fig:dynamic_codebook}. In terms of the directionality of the beam-pattern,  this means that the point on the Earth's surface such that a specific beam  with ID $m$ at time $t$ has its maximum directionality, is also the maximum directionality point for the beam with ID $m$ at time $t+nT_{\rm c}, n\in\mathbb{Z}$.
\begin{figure*}[t!]
    \centering
     \begin{tabular}{ccc}
    \includegraphics[width = 0.3\linewidth, trim={4cm 7.5cm 4cm 6.5cm}, clip]{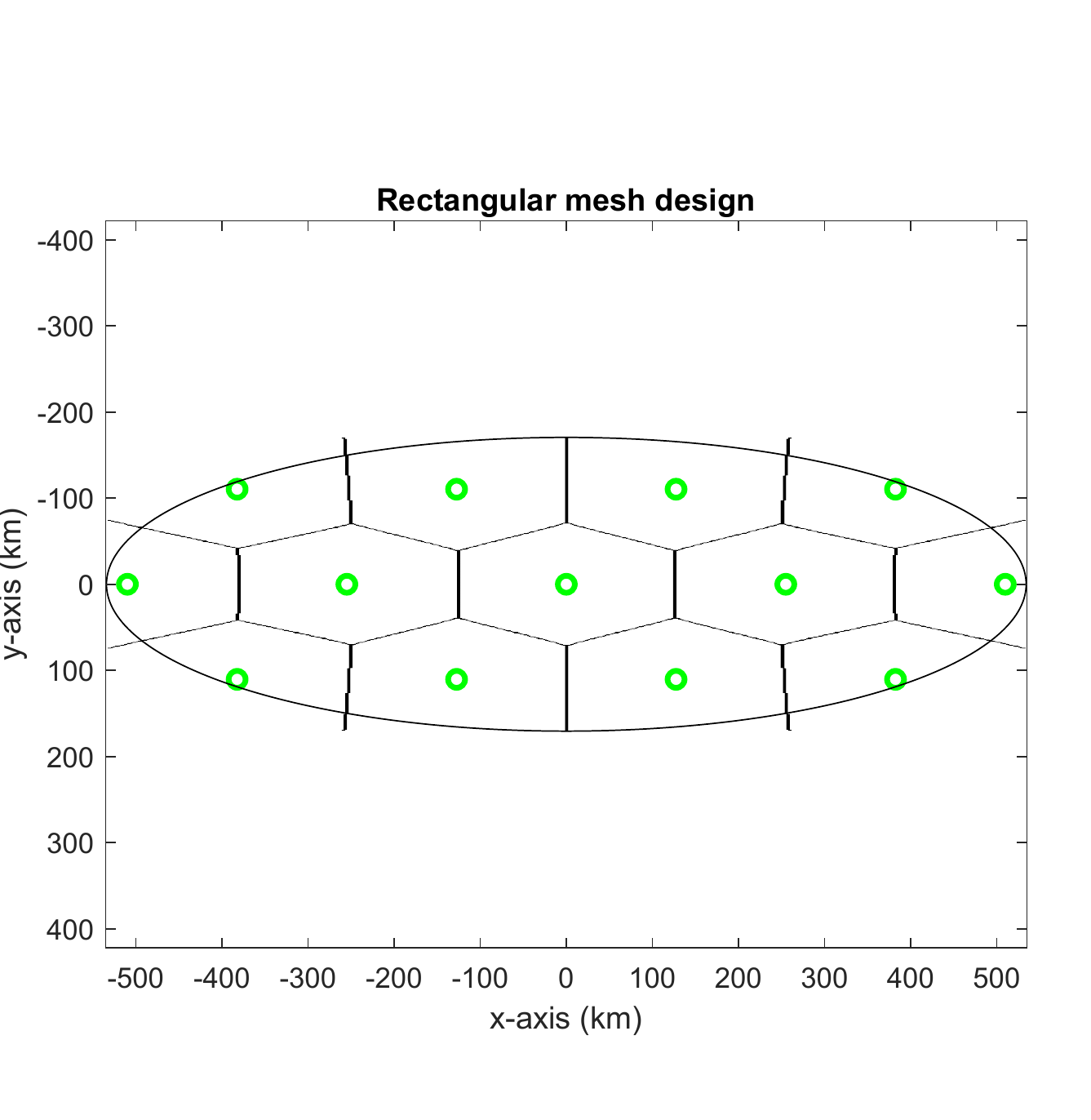}
    & \includegraphics[width = 0.3\linewidth, trim={3.8cm 7.5cm 4cm 6.5cm}, clip]{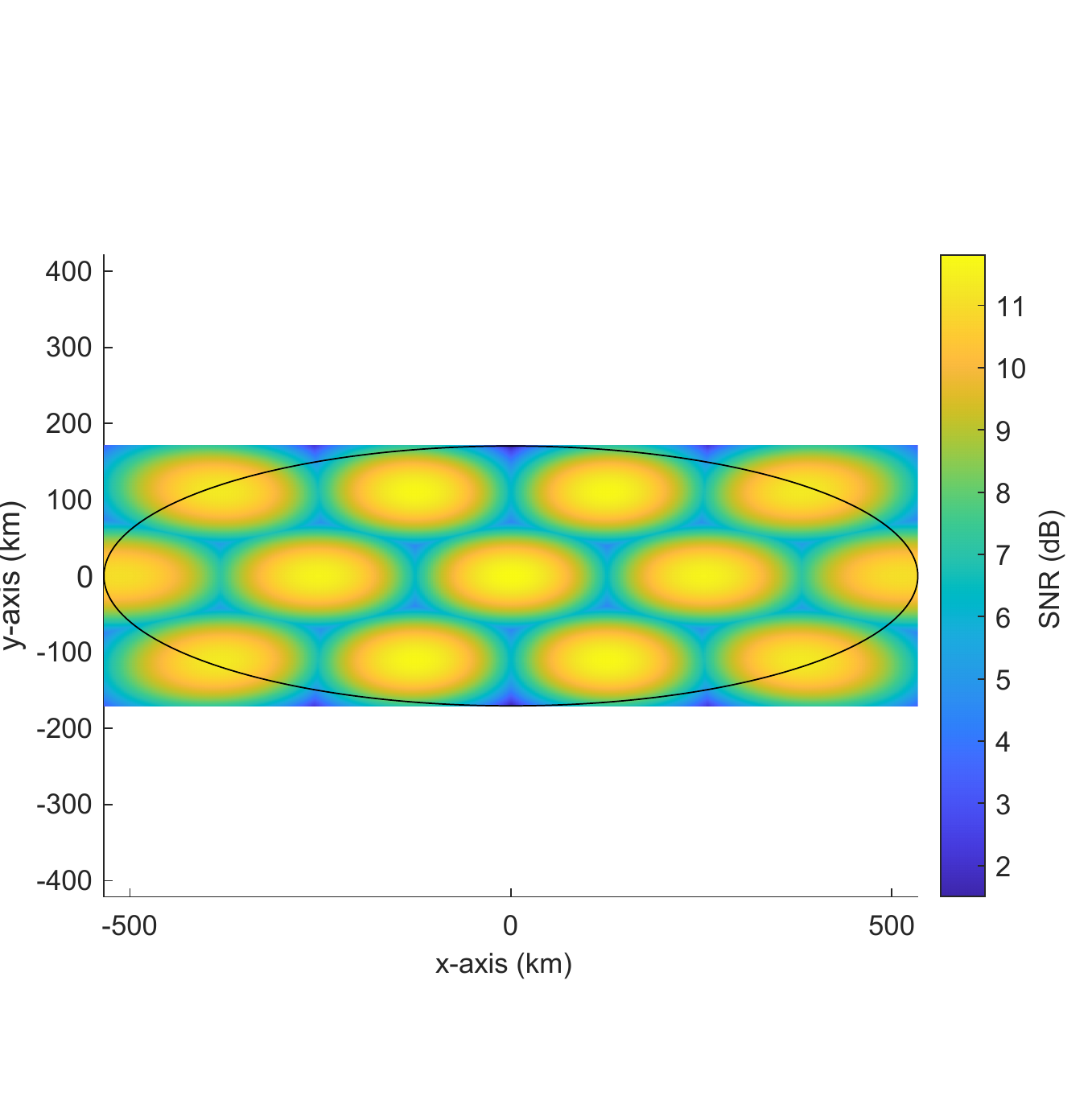}
    & \includegraphics[width = 0.3\linewidth, trim={3.8cm 7.5cm 4cm 6.5cm}, clip]{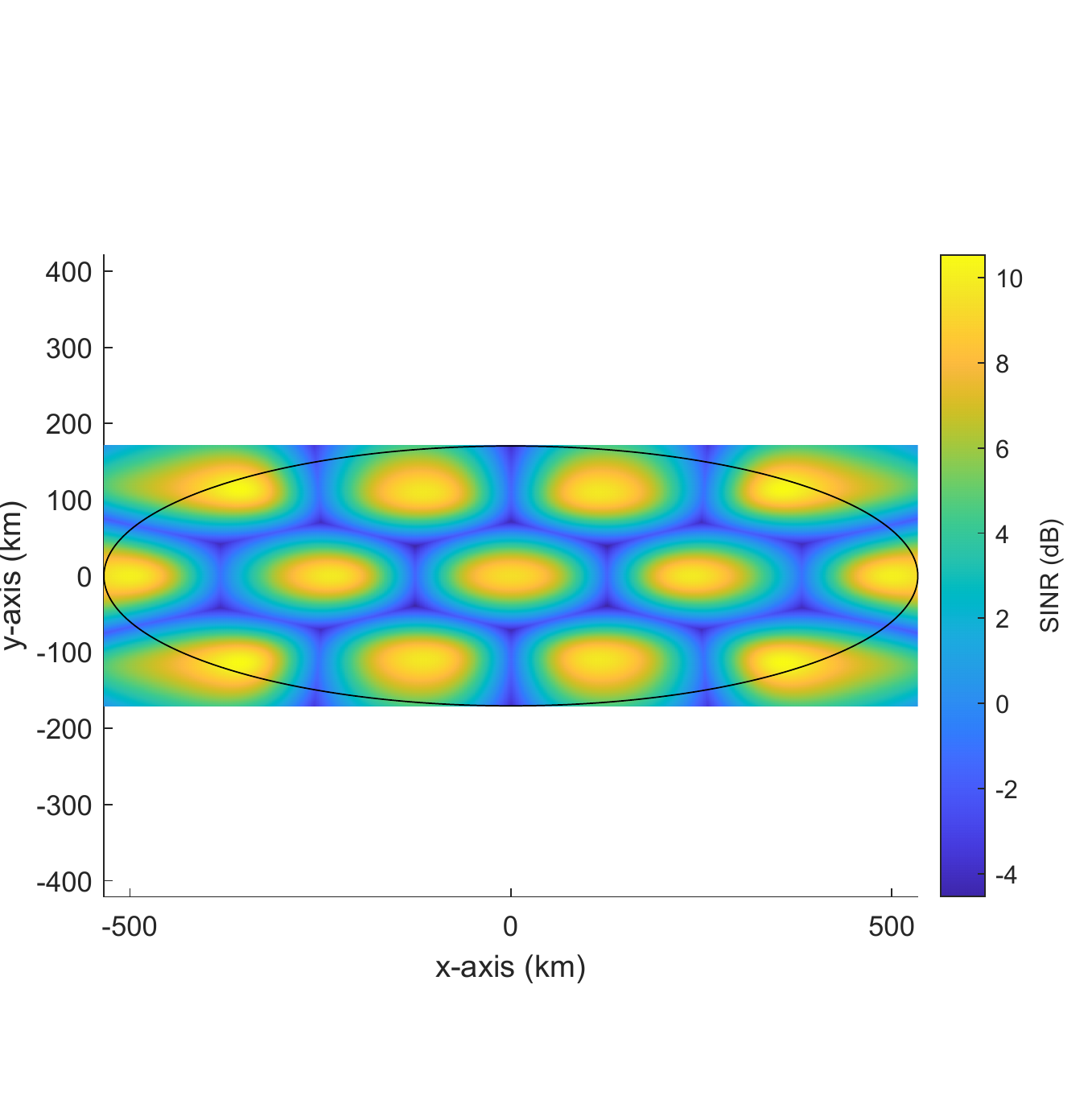}\\
    (a) & (b) & (c) \\
    \end{tabular}
    \caption{(a) Cells associated to the different beams;  points marked in green show the location where the beam gain is maximum. (b) SNR when using the best beam for each cell. (c) SINR when using the best beam for each cell.}
    \label{fig:coverage}
\end{figure*}

After completion of a cycle of the codebook, and due to the satellite movement, the Earth location initially illuminated by the beam  pointing to the lattice point $(x, y)\in\mathcal{L}_k$, becomes illuminated by the beam  pointing to the lattice point $(x-\frac{\pi h_{\rm sat}}{oN^{\rm sub}_{\rm x}}, y)\in\mathcal{L}_k$, so that the beam index changes. To avoid the need of constant handover, a permutation of the beam indices can be applied under the constraint that if $\{(x, y), (x-\frac{\pi h_{\rm sat}}{oN^{\rm sub}_{\rm x}}, y)\}\in\mathcal{L}_k\bigcap\mathcal{E}$, 
then the index of the beam pointing to the lattice point $(x, y)$ gets updated to the index of the beam  pointing to the lattice point $(x-\frac{\pi h_{\rm sat}}{oN^{\rm sub}_{\rm x}}, y)$.
A proposal to address the index updating procedure starts  by determining all the lattice points that will be eventually  active:  $\{(x, y)\in\mathcal{L}_0 \text{ such that }(x-\frac{k\pi h_{\rm sat}}{KoN^{\rm sub}_{\rm x}}, y)\in\mathcal{E}\text{ for }0 \leq k<K\}$, then label  them as  $(x_i,y_i)$ for the  $i$th point, such that $y_i<y_k$ or $x_i<x_j$ if $y_i=y_j$ with respect to the coordinates  $(x_j,y_j)$ of point  $j$ if $i < j$.  With this, the updating procedure  boils down to a cyclic increase of each beam index in one unit.

Note that when using this codebook, the cells on the Earth surface, considered as those locations receiving higher power from a given beam, will be nearly hexagonal, with sides at the equidistant lines to the six nearest lattice points as will be shown in the simulations.  

\begin{table}[h!]
\begin{center}
\begin{tabular}{ |l|c|c|c| }
 \hline \hline
 Parameter & Symbol & Value & Units \\ \hline \hline 
 \multicolumn{4}{|c|}{\em Constellation and ROI} \\ \hline 
 Satellite height  & $h_{\rm sat}$ & $1300$ & km \\
 Number of orbital planes & $N_{\rm p}$ & $83$  & \\
 Number of satellites per orbital plane & $N_{\rm s}$ & $53$ & \\
 Orbital plane inclination & $\theta_\text{op}$ & $53$ & degrees \\ 
 Semiradius of the ROI in $x$-dimension  & $R_{\rm x}$ & 534.1 & km \\
 Semiradius of the ROI in $y$-dimension  & $R_{\rm y}$ & 170.5 & km \\  \hline 
 \multicolumn{4}{|c|}{\em Channel} \\ \hline  
 Carrier frequency & $f_{\rm DL}$ & $11.45$ & GHz \\
 Bandwidth & $B_{\rm DL}$ & $250$ & MHz \\
 Atmospheric path loss & $LP_{\rm at}$ & $0.017$ & dB \\
 Rician factor  & $K_\text{r}$ & $10$ &  \\ \hline
 \multicolumn{4}{|c|}{\em Satellite}  \\ \hline
 Sub-array elements in $x$-dimension & $N^{\rm sub}_{\rm x}$ & $12$ & \\
 Sub-array elements in $y$-dimension  & $N^{\rm sub}_{\rm y}$ & $24$ & \\
 Number of RF chains & $N^{\rm RF}$ & $13$ & \\
 Transmit power & $P_{\rm TX}$ & $15$ & dBW \\
 Oversampling factor  & $o$ & $1.4$  & \\ \hline
 \multicolumn{4}{|c|}{\em User terminal } \\ \hline
User antenna gain & $G_\text{RX}$ & 27.6 & dB\\ 
Receiver noise temperature & T & 24.1 & dB \\\hline
\end{tabular}
\caption{System parameters for the simulation of the downlink of a LEO SatCom system operating in the Ku band.}\label{tab:system_param}
\end{center}
\end{table}

\section{Simulation Results}\label{sec:simulations}
We simulate the downlink of a LEO satellite communication system operating with the parameters described in Table~\ref{tab:system_param}. The LEO constellation includes 53 satellites in each of its 83 orbits, with an orbital plane inclination of $53^\circ$. Each satellite orbits at a  $1300$ km height, covering an ellipsoidal area with semi-radius $534.1$ km in the ${\rm x}$ axis and $170.5$ km in the ${\rm y}$  axis. 
The satellite large phased array consists of $13$ sub-arrays of $12\times24$ antennas, which operate independently.
The gain at the UT corresponds to that of a uniform rectangular array of $24\times 24$ elements, i.e., $G_{\rm RX} = 27.6 {\rm dB}$.

\begin{figure}[!htb]
    \centering
    \includegraphics[width = 0.7\linewidth]{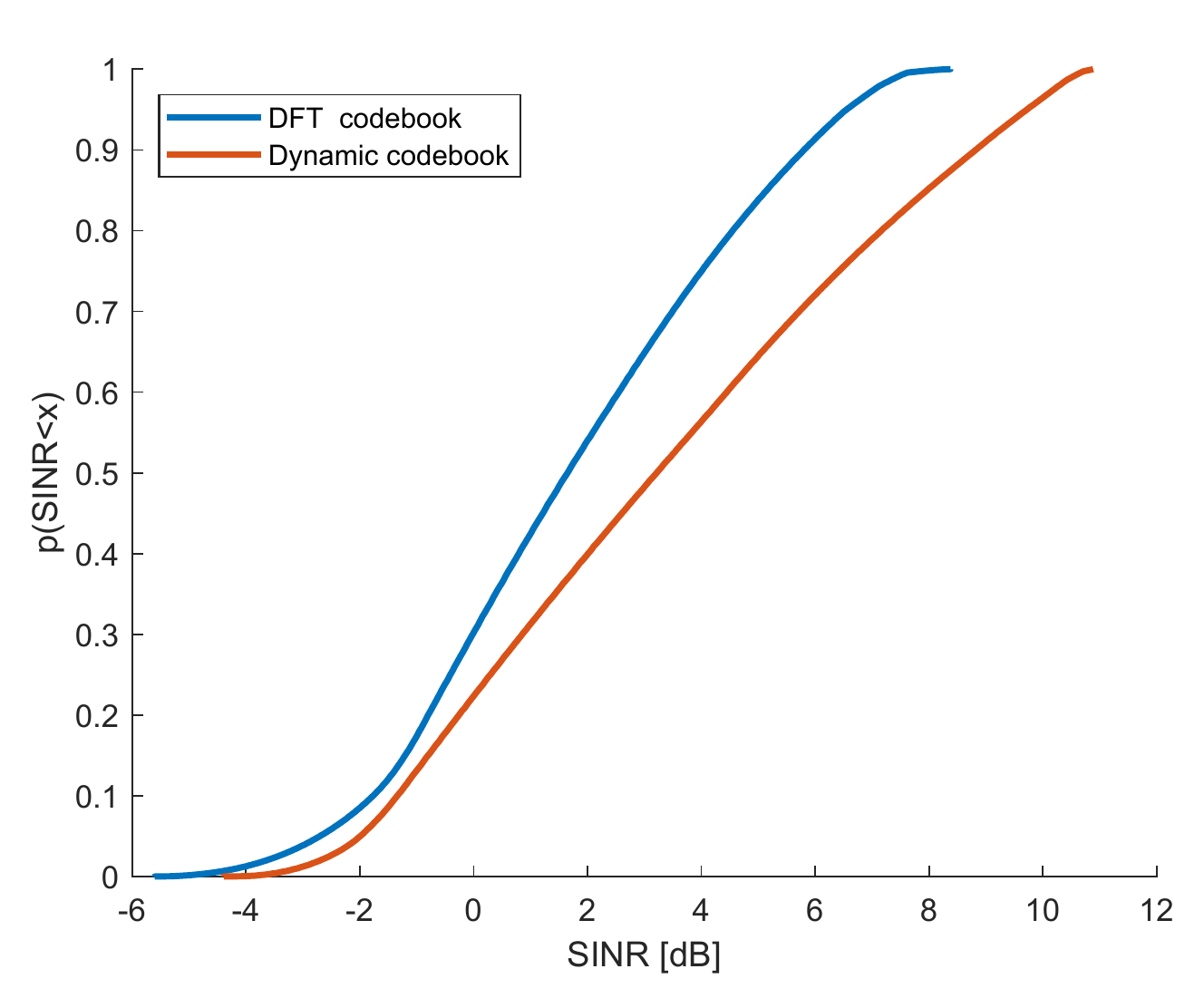}
    \caption{SINR cumulative distribution function for a DFT based approach defined in \cite{PalaciosLEO21} and the proposed dynamic codebook. The hexagonal lattice considered in our initial codebook design introduces a gain in SINR with respect to the rectangular lattice inherent to the 2D-DFT codebook.}
    \label{fig:cdf}
\end{figure}

\begin{figure}[!htb]
    \centering
    \begin{tabular}{cc}
    \includegraphics[width = 0.49\linewidth, trim={4cm 7.5cm 4cm 8.5cm}, clip]{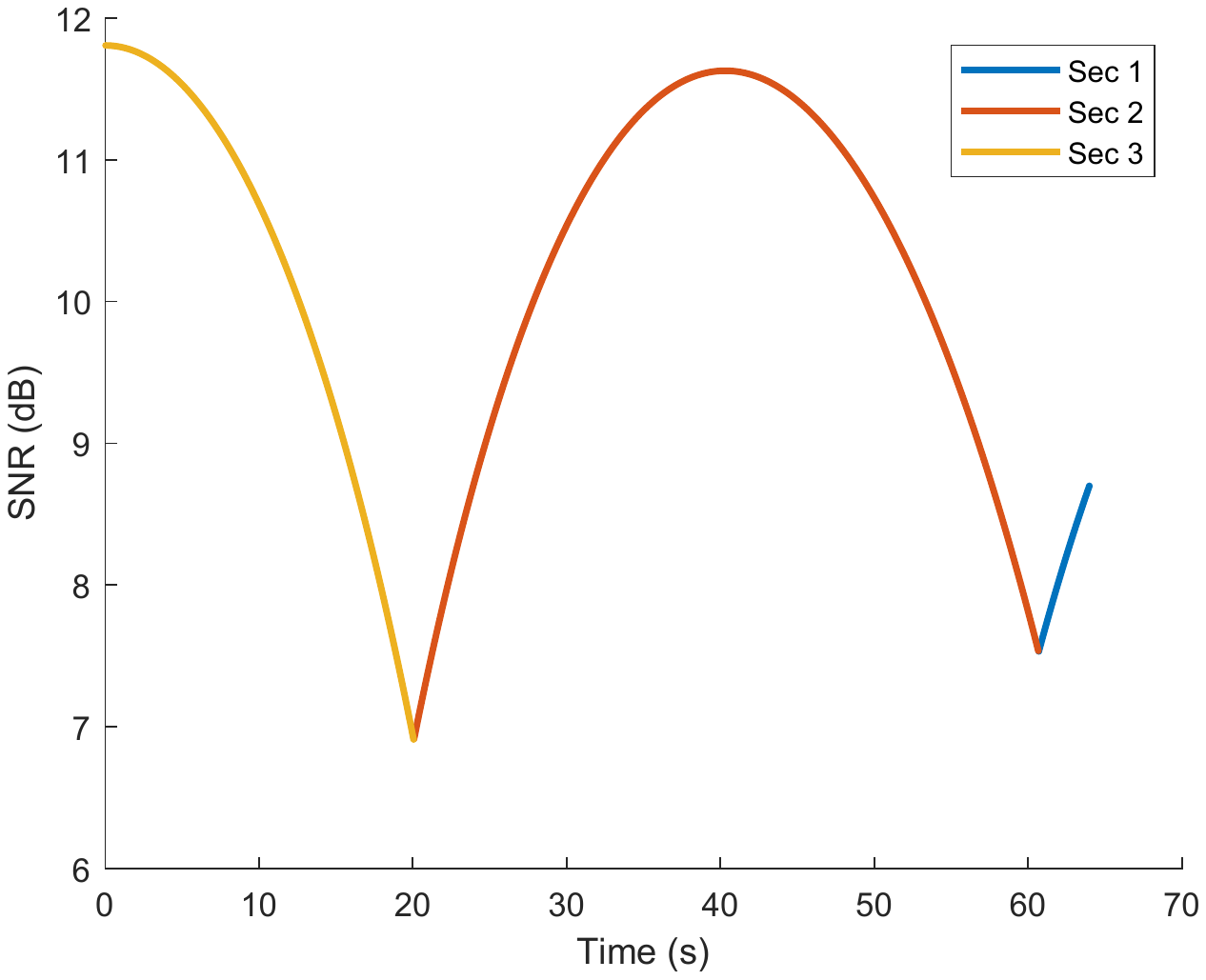} & \includegraphics[width = 0.49\linewidth, trim={4cm 7.5cm 4cm 8.5cm}, clip]{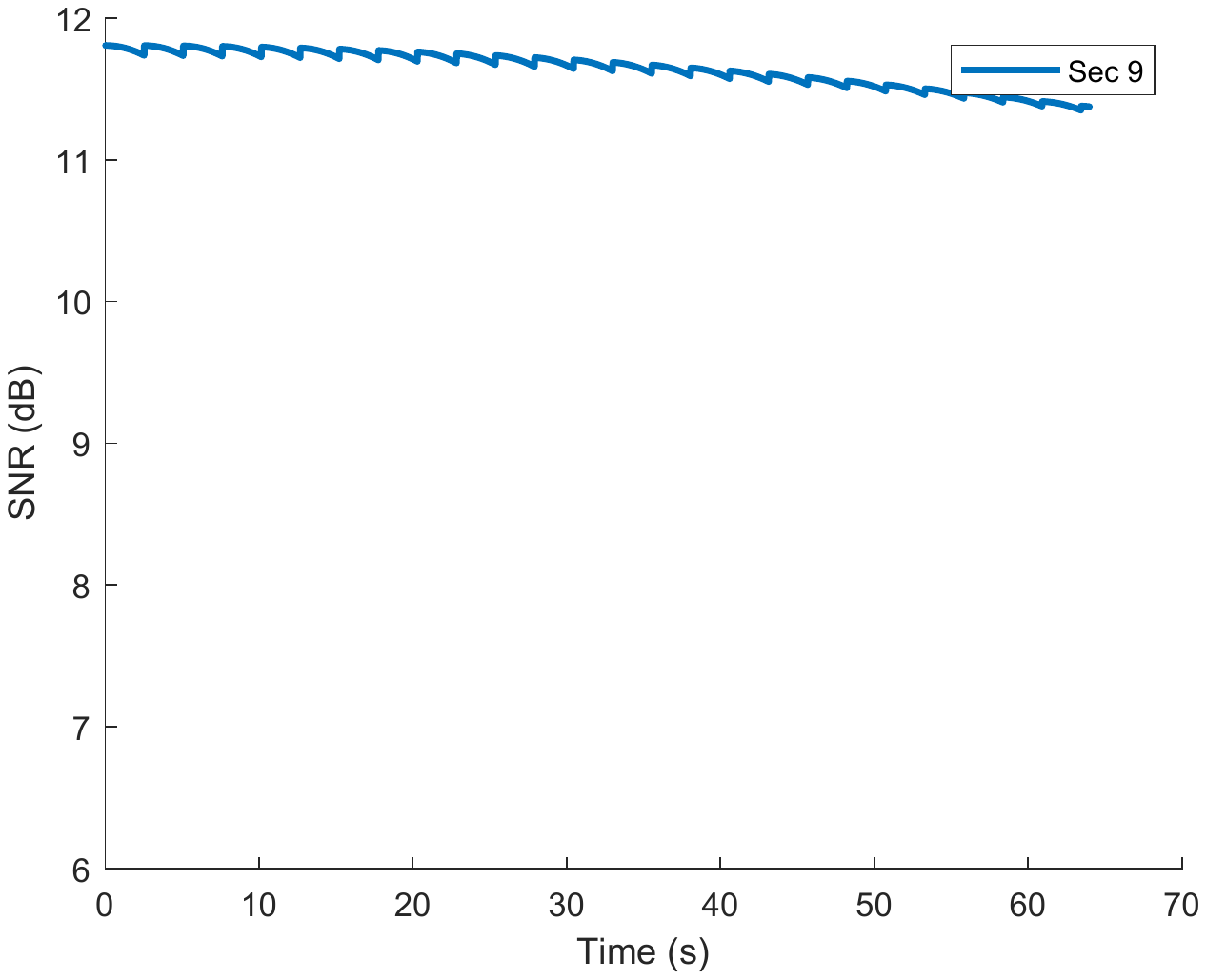}
     \\
     (a) & (b) \\
    \end{tabular}
    \caption{Time SNR variation due to the satellite  movement: (a)  using the DFT codebook in \cite{PalaciosLEO21}; (b) proposed dynamic codebook.}
    \label{fig:degrade}
\end{figure}

\begin{figure}[!htb]
    \centering
    \begin{tabular}{cc}
    \includegraphics[width = 0.49\linewidth, trim={4cm 7.5cm 4cm 8.5cm}, clip]{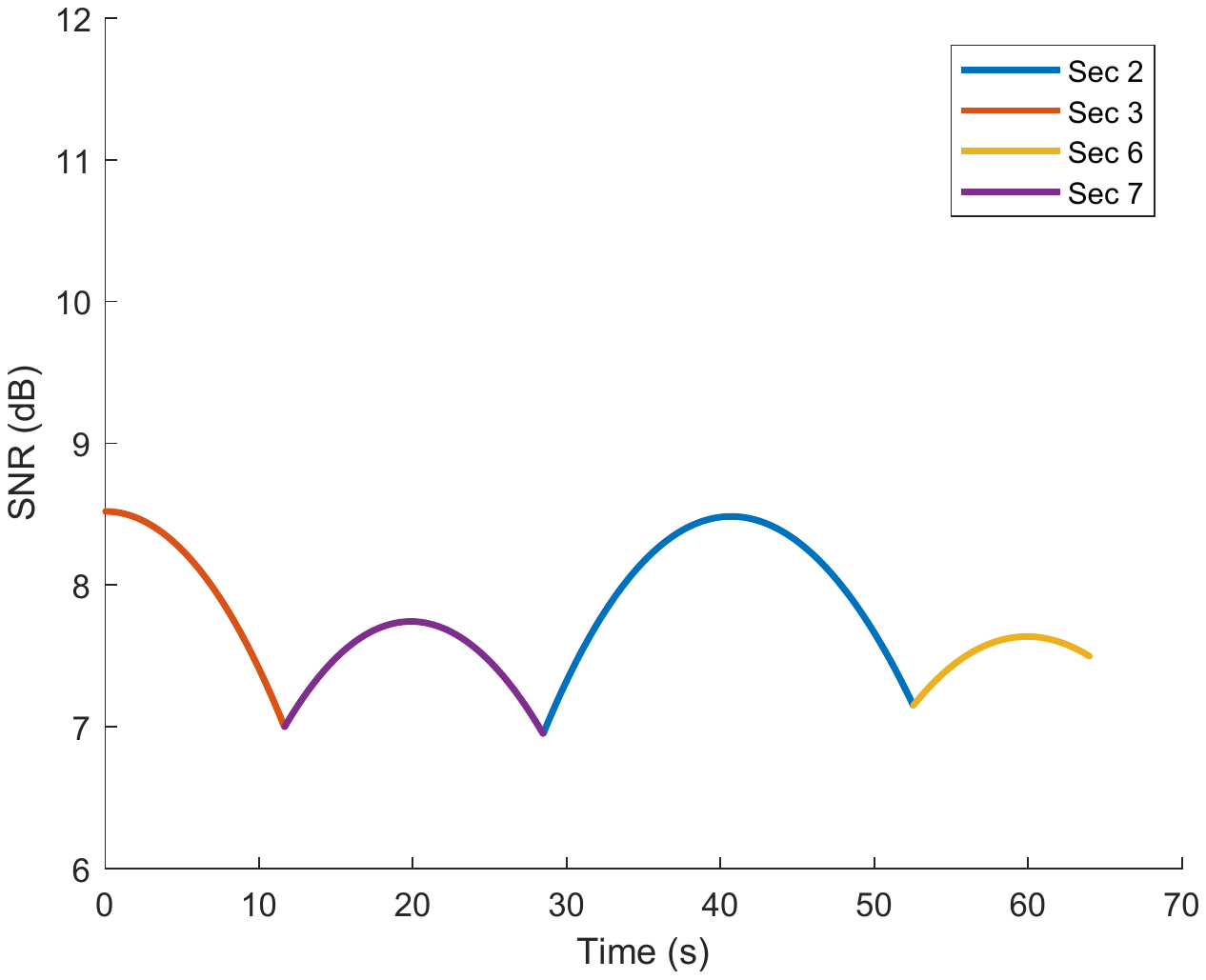} &
     \includegraphics[width = 0.49\linewidth, trim={4cm 7.5cm 4cm 8.5cm}, clip]{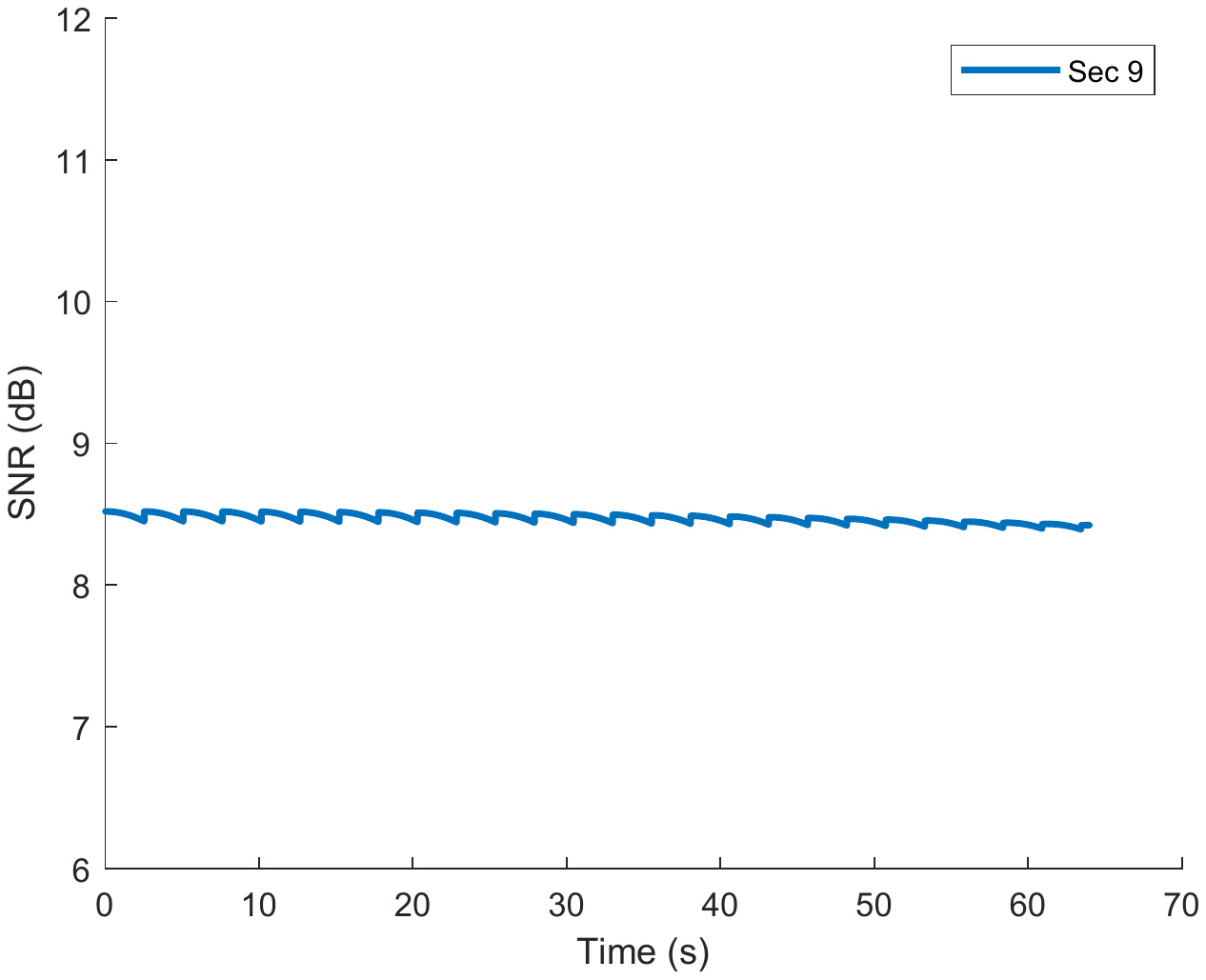}
     \\
     (a) & (b)\\
    \end{tabular}
    \caption{Time SNR variation due to the satellite  movement for a user located 50 km away on the y-axis from the center of the ROP  : (a)  using the DFT codebook in \cite{PalaciosLEO21}; (b) proposed dynamic codebook.}
    \label{fig:degrade_50}
\end{figure}

We start by analyzing the satellite's coverage in the elliptical area when using a static initial codebook as described in Section \ref{sec:initial_codebook}.
To this end, since $LP_{\rm fs}$, $G_{\rm TX}$ and $G_{\rm TX}^{\rm interf}$ are user location dependent, we plot maps of the SNR and SINR in Fig.~\ref{fig:coverage}. The values of the SNR and SINR for each location have been computed using (15) and (19) in \cite{PalaciosLEO21}.
While the SNR values show relatively small variations all over the ellipse, the SINR map shows areas with relative high interference.
This is a consequence of using the same codebook on a single satellite for all users over the whole bandwidth spectrum.
A full frequency reuse leads to SINR values no larger than $-3$  ${\rm dB}$ at those locations covered by three beams with the same gain.
We compare now the performance provided by this initial codebook design with that of an oversampled 2D DFT codebook comprised of 15 beam-patterns, evaluated in \cite{PalaciosLEO21} in the context of a LEO satellite communication system. To that aim, we evaluate the probability of the SINR being larger than a given threshold. Fig. \ref{fig:cdf} shows that 
our design significantly outperforms the 2D DFT codebook used in 5G NR even when using fewer beam-patterns to cover the same area, providing a higher probability of achieving  a given SINR threshold. For example, the probability of having a SINR larger than 4 dB is 0.5 when using the proposed dynamic codebook, and only 0.25  when operating with a fixed 2D DFT codebook. These results show how the initial codebook design proposed in Section \ref{sec:initial_codebook} outperforms the 2D DFT codebook.
Now we evaluate the performance of the dynamic codebook design based on the previous initial codebook. To this aim, we compute the SNR degradation over time due to the satellite's  movement. The UT is initially located at the center of the ROP, which is moving as the satellite travels. Fig.~\ref{fig:degrade} shows the evolution of the SNR value when using the initial codebook without beam adaptation, and the proposed  dynamic codebook. 
When using the initial codebook in a static way, we can see that the user is illuminated for around $30$ seconds by each beam-pattern, with an SNR variation of around 3dB inside the beam footprint. In addition, there is a  progressive degradation of  the peak gain that the best beam can provide. 
With the proposed dynamic codebook, the UT experiences an SNR with a variation of the order of only 0.5 dB over the same period of time.
We consider now the evolution of SNR over time for a user located 50 km away on the y-axis from the center of the ROP, using both the DFT codebook and the dynamic codebook. The results are shown in Fig.~\ref{fig:degrade_50}. In this case, the user experiences a variation of  1.5 dB with the static 2D DFT codebook, while  the dynamic codebook provides approximately the same peak SNR over the analyzed cycles. These two examples show how the dynamic codebook clearly outperforms the static DFT codebook at different user locations.  

\begin{figure}[!htb]
    \centering
    \begin{tabular}{c}
        \vspace*{-2mm}
    \includegraphics[width = 0.75\linewidth, trim={4cm 7.5cm 4cm 8.5cm}, clip]{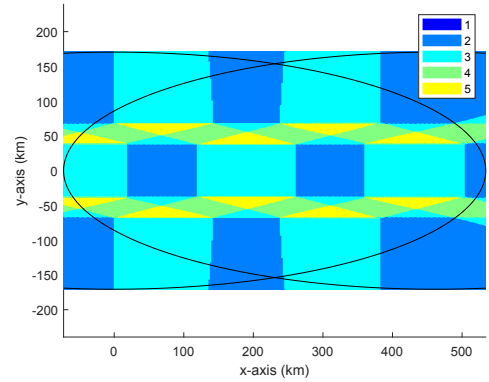} \\
    (a) \\
     \includegraphics[width = 0.75\linewidth, trim={4cm 7.5cm 4cm 8.5cm}, clip]{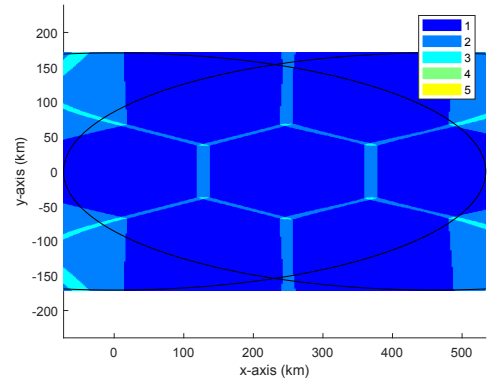}
     \\
     (b) \\
    \end{tabular}
    \caption{Number of handovers experienced by a user at a given location: (a) using the initial codebook in a static way; (b) using the proposed dynamic codebook. }
    \label{fig:handover}
\end{figure}

Finally, we evaluate the ability of the beam ID permutation mechanism to reduce the number of handovers. Fig.~\ref{fig:handover} shows the number of handovers that a user located at a given point would experience when using the static and dynamic codebooks.  It is easy to check that the number of handovers is kept to one for most of the locations in the ROI when using the dynamic codebook, while the static one requires at least three handovers for most of the locations.

\section{Conclusions}
In this paper we have proposed a family of  codebooks specifically designed for analog beamforming in  LEO satellites, which affords some flexibility for beam steering without incurring in extra computational complexity at the payload. The selection of the appropriate codebook from a judiciously designed family allows to offer a smoother coverage to the ground terminals, minimizing handovers while keeping a stable beamforning gain.

\bibliographystyle{IEEEtran}
\bibliography{NTN_NGP}

\end{document}